\documentclass[aps,twocolumn,pra,showpacs,floatfix]{revtex4}
\usepackage{epsfig}
\usepackage{graphicx}
\usepackage{dcolumn}
\usepackage{amsmath}

\begin{document}


\title{Coupled-cluster single-double calculations of the relativistic
energy shifts in C~IV, Na~I, Mg~II, Al~III, Si~IV, Ca~II and Zn~II}

\author{V. A. Dzuba}
\email{V.Dzuba@unsw.edu.au}
\affiliation{School of Physics, University of New South Wales,
Sydney 2052, Australia}

\author{W. R. Johnson}
\email{johnson@nd.edu}
\homepage{www.nd.edu/~johnson}
\affiliation{
Department of Physics, 225 Nieuwland Science Hall\\
University of Notre Dame, Notre Dame, IN 46566}

\date{\today}

\begin{abstract}

The relativistic coupled-cluster single-double method is used to calculate the
dependence of  frequencies of strong $E1$-transitions in many
monovalent atoms and ions on the fine-structure constant $\alpha$.
These transitions are used in the search for manifestations of the variation
of the fine-structure constant in quasar absorption spectra. Results
of the present calculations are in good agreement with previous calculations
but are more accurate.

\end{abstract}


\maketitle

\section{Introduction}

Theories unifying gravity with other interactions suggest a possibility
of temporal and spatial variations of the fundamental  constants of nature;
a review of these theories and results of measurement can be found
in Ref.~\cite{Uzan}.  A very sensitive many-multiplet (MM) method
to search for the variation of the fine-structure constant
$\alpha=e^2/\hbar c$  by comparison of quasar absorption spectra with
laboratory spectra has been suggested in Refs.~\cite{Webb99,Dzuba99}.

Using this method, strong evidence that the fine-structure constant
might be smaller about ten billion years ago was
found~\cite{Webb99,Webb01,Murphy01a,Murphy01b,Murphy01c,Murphy01d}.
This result was obtained from an analysis of data from the Keck
telescope in Hawaii by the group  based at the University
of New South Wales in Australia. However, an analysis of  data
from the VLT telescope in Chile, performed by different groups~\cite{vlt1,vlt2}
using the same MM method, gave a null result. There is an outgoing debate
in the literature about possible reasons for the disagreement.

The MM method requires calculation of  relativistic corrections
to frequencies of atomic transitions to reveal their dependence
on the fine-structure constant.
All calculations used in the analysis so far were performed by a
single research group based at the University of New South
Wales~\cite{Dzuba99a,Dzuba01,Dzuba02,Berengut04,Berengut05,Berengut06,archDzuba}.
Owing to the importance of detecting  variations of fundamental constants and
the disagreement mentioned above,
it is important that the related atomic calculations be verified independently.

A positive development in this direction is a recent independent calculation
of the relativistic energy shifts in the ion Fe~II ~\cite{Porsev}.  Fe~II is
the single most important element in the analysis of quasar absorption
spectra. It has lines which move in opposite directions when $\alpha$ varies;
moreover,  the shifts in Fe~II are relatively large. In principle,
subject to sufficient statistics, Fe~II alone could serve as a probe
of variation of the fine-structure constant in quasar absorption
spectra~\cite{Porsev}. However, calculations for Fe~II are difficult
due to the large number of valence electrons.

In a recent work~\cite{Savukov} independent calculations of
the relativistic energy shifts for monovalent and diavalent atoms
of astrophysical interest were reported. This work
also presented a detailed analysis of Breit contributions to the
relativistic energy shift. Note that,  despite some overlap of authors of
early~\cite{Dzuba99a,Dzuba01,Dzuba02,Berengut04,Berengut05,Berengut06,archDzuba}
and recent~\cite{Porsev,Savukov} works,  the later can be regarded as
independent since they use completely independent sets of computer
codes and different methods of calculation.

In present work, we study the role of higher-order correlations in
relativistic energy shifts of monovalent atoms and ions of
astrophysical interest.  We use a linearized coupled-cluster
method in the single-double approximation and third-order
many-body perturbation theory to perform the calculations. We
demonstrate that including higher-order correlations
significantly improves the accuracy of the transition energies.
The values of relativistic energy shifts found in the present work
are in good agreement with previous calculations but are more
accurate. Note also that the present work can be considered as
an independent verification of earlier calculations.
This is because the calculations are performed using a  method
that has never before been used for this purpose.

Apart from the important task of calculating relativistic energy shifts
of atomic frequencies of astrophysical interest, this work can be
considered as another demonstration of the power of the single-double
method.  It applies the method for the first time
to the ions C~IV and Zn~II.

\section{Method}

It is convenient to present the dependence of atomic frequencies on
the fine-structure constant $\alpha$ in the vicinity of its physical
value $\alpha_0$ in the form
\begin{equation}
  \omega(x) = \omega_0 + qx,
\label{omega}
\end{equation}
where $\omega_0$ is the laboratory value of the frequency and
$x = (\alpha/\alpha_0)^2-1$, $q$ is the coefficient which is to be
found from atomic calculations. Note that
\[ q = \left .\frac{d\omega}{dx}\right|_{x=0}.\]
To calculate this derivative numerically we use
\begin{equation}
  q \approx  \frac{\omega(+\delta) - \omega(-\delta)}{2\delta}.
\label{deriv}
\end{equation}
Here $\delta$ must be small to exclude non-linear in $\alpha^2$ terms.
In the present calculations we use $\delta = 0.05$, which leads to
\begin{equation}
  q \approx  10 \left(\omega(+0.05) - \omega(-0.05)\right).
\label{deriv05}
\end{equation}
To calculate the coefficients $q$ using (\ref{deriv05}),  $\alpha$ must be varied
in our computer codes.  Therefore, it is convenient to use a form
of the single electron wave function in which the dependence on $\alpha$ is
explicitly shown (we use atomic units in which $e=\hbar=1, \alpha = 1/c$)
\begin{equation}
    \psi(r)_{njlm}=\frac{1}{r}\left(\begin {array}{c}
    f_{v}(r)\Omega(\mathbf{n})_{\mathit{jlm}}  \\[0.2ex]
    i\alpha g_{v}(r)  \widetilde{ \Omega}(\mathbf{n})_{\mathit{jlm}}
    \end{array} \right),
\label{psi}
\end{equation}
where $n$ is the principal quantum number and an index $v$
replaces the three-number set $n,j,l$.
This leads to a form of radial equation for single-electron
orbitals which also explicitly depends on $\alpha$:
\begin{equation}
    \begin {array}{c} \dfrac{df_v}{dr}+\dfrac{\kappa_{v}}{r}f_v(r)-
    \left[2+\alpha^{2}(\epsilon_{v}-\hat{V}_{HF})\right]g_v(r)=0,  \\[0.5ex]
    \dfrac{dg_v}{dr}-\dfrac{\kappa_{v}}{r}f_v(r)+(\epsilon_{v}-
    \hat{V}_{HF})f_v(r)=0, \end{array}
\label{Dirac}
\end{equation}
here $\kappa=(-1)^{l+j+1/2}(j+1/2)$,
and $\hat{V}_{HF}$ is the Hartree-Fock potential.
Equation (\ref{Dirac}) with $\alpha = \alpha_0 \sqrt{\delta +1}$
is used to construct a full set of single-electron orbitals.

As a first step, equation (\ref{Dirac}) is used to calculate self-consistently
single-electron states of the closed-shell core. Then, this equation
is used to calculate a complete set of B-spline single-electron basis
orbitals. We use 40 B-splines of order $k=9$ in a cavity of radius
40$a_B$ with angular momentum $l$ up to $l_{max}=5$.
More details on the use of B-splines
in atomic calculations can be found in Ref.~\cite{Bsplines}.

This basis set is used to perform  calculations with the
linearized couple-cluster single-double
method supplemented by  third-order many-body perturbation theory.
The all-order single-double (SD) method was discussed before in
Refs.~\cite{sd1,sd2,sd3,sd4,sd5} and third-order many-body perturbation
theory was discussed in Refs.~\cite{sd3,mbpt3}.
We also include results from second-order  MBPT to uncover the role
of higher-order correlations and to estimate the numerical uncertainty
due to correlations.

The SD equations are written for the coefficients of the
expansion of the many-electron wave function in terms of single
and double excitations from the reference Hartree-Fock wave
function. They contain Coulomb integrals between single-electron
basis orbitals as parameters. The equations are first solved iteratively
for the atomic core. When convergence is achieved, a similar procedure
for valence states of interest is  performed.

Many-body perturbation theory is used to calculate the third-order
diagrams missed in the SD method (E3$_{\text{extra}}$). Therefore,
the present
calculations are complete through  third order while selected classes of
higher-order diagrams are included in all orders.  A number of
earlier calculations (see, e.g.\ \cite{sd1,sd2,sd3,sd4,sd5,mbpt3})
prove that this approach gives very good accuracy for a wide
range of monovalent atoms and ions.

The SD equations depend on the fine-structure constant only implicitly,
via the values of the Coulomb integrals, which are affected by the change
of the single-electron basis states (\ref{psi}) due to change of $\alpha$
in the Hamiltonian (\ref{Dirac}). In other words,
changes in $\alpha$ lead to changes in the basis orbitals, while the SD
equations remain unchanged. The same is true for terms in the MBPT
expansion.
However, the change of basis means that all SD and E3 calculations must
be redone. The E3 calculations must be repeated from scratch, but the SD
iterations can be restarted from previous calculations for a different
value of $\alpha$.  Since the change of $\alpha$ is small,
[see Eq.~(\ref{deriv})] only a few iterations are needed to solve
the SD equations for new value of $\alpha$.
 This significantly speeds up the calculations.

Although the method used in present work has been used many times before,
the set of computer codes used in present work is new and independent
of previous versions. The main reason for developing new codes was
the need to have them in a form that allows easy modifications, as
e.g. the change of the fine-structure constant in this calculations, or
inserting extra operators (Breit interaction, specific mass shift
operator, etc.)  for future projects.
Some features of the present realization of the methods
are presented in the appendix.  For third-order MBPT calculations,
we apply a universal algorithm that uses a numerical description of
the E3 MBPT diagrams and the same piece of computer code to calculate
all of them.
This algorithm will be described in detail elsewhere \cite{Dzuba08}.

\section{Results}

\begin{table*}
\caption{Removal energies of the lowest $s$ and $p$ states of
C~IV, Na~I, Mg~II, Al~III, Si~IV, Ca~II and Zn~II in different
approximations, comparison with experiment (cm$^{-1}$).}
\label{energies}
\begin{ruledtabular}
\begin{tabular}{lclrrrrrrrr}
\multicolumn{1}{c}{Atom} & \multicolumn{1}{c}{$Z$} &
\multicolumn{1}{c}{State} & \multicolumn{1}{c}{RHF} &
\multicolumn{1}{c}{E2} & \multicolumn{1}{c}{$\Delta$\footnotemark[1]} &
\multicolumn{1}{c}{SD} &  \multicolumn{1}{c}{$\Delta$\footnotemark[1]} &
\multicolumn{1}{c}{SD+E3} &  \multicolumn{1}{c}{$\Delta$\footnotemark[1]} &
\multicolumn{1}{c}{Expt\footnotemark[2]} \\
\hline
C~IV   & 6  & 2$s$       & 519255 & 520189 &    11 & 520233 &  55  & 520231 &  53 & 520178 \\
       &    & 2$p_{1/2}$ & 454054 & 455640 &   -54 & 455732 &  38  & 455727 &  33 & 455694 \\
       &    & 2$p_{3/2}$ & 453927 & 455509 &   -78 & 455600 &  13  & 455595 &   8 & 455587 \\

Na~I   & 11 & 3$s$       &  39952 &  41229 &  -220 &  41437 & -12  &  41450 &   1 &  41449 \\
       &    & 3$p_{1/2}$ &  24030 &  24412 &   -81 &  24486 &  -7  &  24489 &  -4 &  24493 \\
       &    & 3$p_{3/2}$ &  24014 &  24394 &   -82 &  24468 &  -8  &  24471 &  -5 &  24476 \\

Mg~II  & 12 & 3$s$       & 118824 & 121076 & -192  & 121278 &  10  & 121273 &  5  & 121268 \\
       &    & 3$p_{1/2}$ &  84294 &  85453 & -145  &  85585 & -13  &  85586 & -12 &  85598 \\
       &    & 3$p_{3/2}$ &  84204 &  85357 & -150  &  85488 & -19  &  85489 & -18 &  85507 \\

Al~III & 13 & 3$s$       & 226396 & 229319 & -127  & 229489 &  43  & 229464 &  18 & 229446 \\
       &    & 3$p_{1/2}$ & 173687 & 175608 & -155  & 175751 & -12  & 175741 & -22 & 175763 \\
       &    & 3$p_{3/2}$ & 173452 & 175361 & -168  & 175504 & -25  & 175494 & -35 & 175529 \\

Si~IV  & 14 & 3$s$       & 360614 & 364033 &   -60 & 364172 &   79 & 364132 &  39 & 364093 \\
       &    & 3$p_{1/2}$ & 290074 & 292667 &  -139 & 292802 &   -4 & 292780 & -26 & 292806 \\
       &    & 3$p_{3/2}$ & 289606 & 292183 &  -161 & 292317 &  -27 & 292296 & -48 & 292344 \\

Ca~II  & 20 & 4$s$       &  91440 &  96173 &   425 &  96097 &  349 &  95577 &-171 &  95748 \\
       &    & 4$p_{1/2}$ &  68037 &  70680 &   123 &  70761 &  204 &  70491 & -66 &  70557 \\
       &    & 4$p_{3/2}$ &  67837 &  70449 &   115 &  70529 &  195 &  70262 & -72 &  70334 \\

Zn~II  & 30 & 4$s$       & 135134 & 143835 & -1055 & 144618 & -272 & 145334 & 444 & 144890 \\
       &    & 4$p_{1/2}$ &  90524 &  95249 & -1161 &  96184 & -226 &  96613 & 203 &  96410 \\
       &    & 4$p_{3/2}$ &  89787 &  94372 & -1164 &  95311 & -225 &  95728 & 192 &  95536 \\
\end{tabular}
\end{ruledtabular}
\noindent \footnotetext[1]{$\Delta = E_{\rm calc} - E_{\rm expt}$}
\noindent \footnotetext[2]{NIST, Ref.~\cite{NIST}}
\end{table*}

Results of calculations of energy levels of C~IV, Na~I, Mg~II, Al~III,
Si~IV, Ca~II and Zn~II are presented in Table~\ref{energies}.
Removal energies of the lowest $s$ and $p$-states, which are important
for the analysis of quasar absorption data, are presented in
different approximations.
These include the relativistic Hartree-Fock approximation (RHF),
second-order many body perturbation theory (E2), single-double (SD)
approximation, and SD supplemented by  third-order many-body
perturbation theory (SD+E3).  For each approximation the difference
between theoretical and experimental energies is presented in the
columns headed $\Delta$.

The Mg~II, Al~III, and Si~IV ions represent an isoelectronic sequence
of sodium. This sequence was considered in detail in Ref.~\cite{sd3}.
The results of present work are in good agreement with previous calculations.
Some small difference can be attributed to the difference in numerical
procedures and numerical parameters (such as the number of splines, the cavity radius,
maximum angular momentum, etc.)

The results in Table~\ref{energies} show that correlations are large
and are strongly dominated by second-order MBPT. However, inclusion
of higher-order correlations is important and leads to
a significant reduction in the differences between theoretical
and experimental energies.
\begin{table}
\caption{Relativistic energy shifts ($q$-coefficients, see Eq.~(\ref{omega}) for
the lowest $s$ and $p$ states of
C~IV, Na~I, Mg~II, Al~III, Si~IV, Ca~II and Zn~II in different
approximations (cm$^{-1}$).}
\label{qp}
\begin{ruledtabular}
\begin{tabular}{lclrrrr}
\multicolumn{1}{c}{Atom} & \multicolumn{1}{c}{$Z$} &
\multicolumn{1}{c}{State} & \multicolumn{1}{c}{RHF} &
\multicolumn{1}{c}{E2} & \multicolumn{1}{c}{SD} &
\multicolumn{1}{c}{SD+E3} \\
\hline
C~IV   & 6  & 2$s$       &  243   &  244   &   244  &  244  \\
       &    & 2$p_{1/2}$ &  139   &  141   &   142  &  142  \\
       &    & 2$p_{3/2}$ &   11   &   10   &    10  &   10  \\

Na~I   & 11 & 3$s$       &   51   &   57   &    59  &   59  \\
       &    & 3$p_{1/2}$ &   12   &   13   &    13  &   13  \\
       &    & 3$p_{3/2}$ &   -5   &   -5   &    -6  &   -5  \\

Mg~II  & 12 & 3$s$       &  181   &  193   &   194  &  194  \\
       &    & 3$p_{1/2}$ &   70   &   74   &    74  &   74  \\
       &    & 3$p_{3/2}$ &  -20   &  -23   &   -23  &  -23  \\

Al~III & 13 & 3$s$       &  405   &  421   &   422  &  422  \\
       &    & 3$p_{1/2}$ &  197   &  203   &   203  &  203  \\ 
       &    & 3$p_{3/2}$ &  -39   &  -45   &   -45  &  -44  \\ 

Si~IV  & 14 & 3$s$       &  753   &  773   &   774  &  774  \\
       &    & 3$p_{1/2}$ &  416   &  425   &   425  &  425  \\ 
       &    & 3$p_{3/2}$ &  -54   &  -62   &   -62  &  -61  \\ 

Ca~II  & 20 & 4$s$       &  354   &  396   &   394  &  392  \\
       &    & 4$p_{1/2}$ &  161   &  176   &   176  &  175  \\ 
       &    & 4$p_{3/2}$ &  -41   &  -58   &   -59  &  -56  \\ 

Zn~II  & 30 & 4$s$       & 2352   & 2863   &  2873  & 2910  \\
       &    & 4$p_{1/2}$ &  990   & 1271   &  1333  & 1359  \\ 
       &    & 4$p_{3/2}$ &  224   &  354   &   412  &  427  \\ 
\end{tabular}
\end{ruledtabular}
\end{table}

Table~\ref{qp} presents relativistic energy shifts ($q$ coefficients)
in the same approximations as the energies in Table~\ref{energies}.
One can see that the role of correlations is much less important
for the $q$-coefficients than for energies. While second-order
correlations give some rather small contribution to $q$,
contributions from higher-order correlations are practically negligible
in most cases. Only for Zn~II the higher-order contributions are
significantly larger than the uncertainty of the calculations.
Thus,  calculations of the $q$-coefficients are more stable
than the calculations of energies and, consequently, more accurate.

Breit corrections to the frequencies of the transitions considered in present paper
were calculated in Ref.~\cite{Savukov}. These corrections are larger than the uncertainty
of the $q$-coefficients from omitted higher-order correlations. Therefore,
they must be included for accurate results.  Table~\ref{qs}
summarizes all significant contributions. The Dirac contributions
in this Table are based on averaging the SD and SD+E3 approximations
from Table~\ref{qp}.  Breit corrections are taken from Ref.~\cite{Savukov}.

\begin{table}
\caption{Contributions to the relativistic energy shifts
($q$-coefficients, see Eq.~(\ref{omega}) for
the  $s$ - $p$ transitions in
C~IV, Na~I, Mg~II, Al~III, Si~IV, Ca~II and Zn~II  (cm$^{-1}$)}
\label{qs}
\begin{ruledtabular}
\begin{tabular}{lclrrr}
\multicolumn{1}{c}{Atom} & \multicolumn{1}{c}{$Z$} &
\multicolumn{1}{c}{Transition} &
\multicolumn{1}{c}{Dirac\footnotemark[1]} &
\multicolumn{1}{c}{Breit\footnotemark[2]} &
\multicolumn{1}{c}{Total} \\
\hline
C~IV   & 6  & $2s- 2p_{1/2}$ &  102  &   13    &  115 \\
       &    & $2s- 2p_{3/2}$ &  234  &  -12    &  222 \\
Na~I   & 11 & $3s- 3p_{1/2}$ &   46  &   -1    &   45 \\
       &    & $3s- 3p_{3/2}$ &   64  &   -2    &   62 \\
Mg~II  & 12 & $3s- 3p_{1/2}$ &  120  &    1    &  121 \\
       &    & $3s- 3p_{3/2}$ &  217  &   -5    &  212 \\
Al~III & 13 & $3s- 3p_{1/2}$ &  219  &    5    &  224 \\
       &    & $3s- 3p_{3/2}$ &  467  &   -9    &  458 \\
Si~IV  & 14 & $3s- 3p_{1/2}$ &  348  &   13    &  361 \\
       &    & $3s- 3p_{3/2}$ &  835  &  -12    &  823 \\
Ca~II  & 20 & $4s- 4p_{1/2}$ &  218  &    3    &  222 \\
       &    & $4s- 4p_{3/2}$ &  450  &   -4    &  446 \\
Zn~II  & 30 & $4s- 4p_{1/2}$ & 1546  &   -5    & 1541 \\
       &    & $4s- 4p_{3/2}$ & 2472  &  -20    & 2452 \\
\end{tabular}
\end{ruledtabular}
\noindent \footnotetext[1]{This work, see Table~\ref{qp}}
\noindent \footnotetext[2]{Savukov and Dzuba, Ref.~\cite{Savukov}}
\end{table}
Our final results are presented in Table~\ref{q-t}.  An estimate of  the numerical
uncertainty is also given and  results are compared with previous calculations.
There are two sources of numerical uncertainty. One is omission of certain higher-order
correlations and numerical accuracy of the SD and SD+E3 calculations. This uncertainty
was estimated by comparing the results in the SD and SD+E3 approximations.
Another source of uncertainty is the accuracy of calculation of Breit contribution.
Breit contributions are calculated very accurately within the relativistic
Hartree-Fock approximation (see Ref.~\cite{Savukov} for details). The only
uncertainty which may come from the  Breit interaction is due to the fact that
correlation corrections to the Breit interaction are ignored.
However, these contributions are small for the relatively light atoms considered
in present paper.
A strong argument that the Breit contribution is calculated sufficiently accurately
was presented in Ref.~\cite{Savukov}, where it was demonstrated that the inclusion
of the Breit interaction brings the theoretical fine-structure intervals into perfect
agreement with experiment. We use a rather conservative estimate of 20\%
for the accuracy of the calculations of Breit contributions.

Table~\ref{q-t} also presents results from previous calculations. All
previous calculations were done in second-order of MBPT. The
work of Ref.~\cite{Savukov} also included Breit corrections.
Results of the present paper are in excellent agreement with this work.
Some difference for Zn~II is due to higher-order correlations.
The results are also in good agreement with early calculations
compiled together in Ref.~\cite{archDzuba}. The principal sources of the small
differences between the  present results and those of Ref.~\cite{archDzuba}
are  Breit and higher-order correlation corrections.

\begin{table}
\caption{Relativistic energy shifts ($q$-coefficients, see Eq.~(\ref{omega}) for
the  $s$ - $p$ transitions in
C~IV, Na~I, Mg~II, Al~III, Si~IV, Ca~II and Zn~II  (cm$^{-1}$);
comparison with other calculations.}
\label{q-t}
\begin{ruledtabular}
\begin{tabular}{lclrrrr}
\multicolumn{1}{c}{Atom} & \multicolumn{1}{c}{$Z$} &
\multicolumn{1}{c}{Transition} & \multicolumn{1}{c}{This} &
\multicolumn{2}{c}{Other} \\
 & & & \multicolumn{1}{c}{work} & \multicolumn{1}{c}{Savukov\footnotemark[1]} &
\multicolumn{1}{c}{Berengut\footnotemark[2]}  \\
\hline \hline
C~IV   & 6  & $2s- 2p_{1/2}$ &  115(2) &  115 & 104(20)  \\
       &    & $2s- 2p_{3/2}$ &  222(2) &  221 & 232(20)  \\
Na~I   & 11 & $3s- 3p_{1/2}$ &   45(0) &   44 & 45(4)    \\
       &    & $3s- 3p_{3/2}$ &   62(0) &   61 & 63(4)    \\
Mg~II  & 12 & $3s- 3p_{1/2}$ &  121(1) &  120 &120(10)  \\
       &    & $3s- 3p_{3/2}$ &  212(1) &  211 &211(10)  \\
Al~III & 13 & $3s- 3p_{1/2}$ &  224(1) &  223 &216(14)  \\
       &    & $3s- 3p_{3/2}$ &  458(2) &  457 &464(30)  \\
Si~IV  & 14 & $3s- 3p_{1/2}$ &  361(2) &  360 &   346   \\
       &    & $3s- 3p_{3/2}$ &  823(2) &  823 &   862    \\
Ca~II  & 20 & $4s- 4p_{1/2}$ &  222(1) &  222 &   224   \\
       &    & $4s- 4p_{3/2}$ &  446(3) &  450 &   452   \\
Zn~II  & 30 & $4s- 4p_{1/2}$ & 1541(7) & 1585 &  1584(25)  \\
       &    & $4s- 4p_{3/2}$ & 2452(13)& 2488 &  2479(25)  \\
\end{tabular}
\end{ruledtabular}
\noindent \footnotetext[1]{Second-order+Breit, Ref.~\cite{Savukov}}
\noindent \footnotetext[1]{Second-order, compilation of previous results
presented in Ref.~\cite{archDzuba}}
\end{table}

\section{Conclusions}

In the present work, we studied the role of higher-order correlations on
relativistic energy shifts of atomic frequencies used in the search for
variations of the fine-structure constant in quasar absorption spectra.
We have demonstrated that the higher-order correlations are important for energies,
bringing theoretical values into better agreement with experiment.  However,
higher-order correlations give very small contributions to the relativistic
energy shifts ($q$-coefficients) in all cases except Zn~II.
Results of the present work are in good agreement
with previous calculations but are more accurate.

\section*{Acknowledgments}

An important part of this work was done during V.A.D. visit to the
University of Notre Dame. He would likes to thank the Department
of Physics of this university for the hospitality and support.
The work was partly supported by Australia Research Council.
The work of W.R.J. was supported in part by NSF Grant No. PHY-04-56828.
\appendix

\section{Efficient way of calculating the SD and MBPT terms}

The coupled-cluster SD method combined with MBPT has led to an
accurate description of many properties of monovalent atoms; however,
 it is very demanding computationally.  For example, for Zn~II the
total number of single-electron basis states used in present
calculations is 319. The total number of non-zero distinctive
Coulomb integrals is about $1.7 \times 10^8$. All of them are
used in both, the SD equations and the MBPT expansion.
Calculating all Coulomb integrals in advance
and keeping them in computer memory is practically impossible
due to huge demand for computer memory. On the
other hand, calculation of Coulomb integrals from single-electron
basis functions every time they are needed makes the calculations
unacceptably slow. This is even more so in the case
of the relativistic energy shifts considered in present paper,
since we need to run all relevant codes several times
for several different values of the fine-structure constant $\alpha$.

To improve the efficiency of the codes, we use an approach in which Hartree
screening functions $Y$ rather than Coulomb integrals are calculated in advance
and kept in memory for efficient calculation of Coulomb integrals
(a similar approach was used in Ref.~\cite{core}).
The Hartree screening function $Y$ is defined as
\begin{multline}
  Y_{knm}(r) = \int \dfrac{r_<^k}{r_>^{k+1}}(f_n(r')f_m(r')   \\
 +  \alpha^2 g_n(r')g_m(r'))dr',
\label{Yknm}
\end{multline}
where $r_< = {\rm min}(r,r')$ and $r_> = {\rm max}(r,r')$.
We also need the $\rho$ functions:
\begin{equation}
  \rho_{jl}(r) = f_j(r)f_l(r)+\alpha^2 g_j(r)g_l(r).
\label{rho}
\end{equation}

Our typical coordinate grid consists of about 1000 points.  Usually all
points are used to calculate $Y_{knm}(r)$. However, there is
no need to keep all points  for successive calculations of the Coulomb integrals.
 It turns out that very little lose of accuracy is caused by
using a subset of points defined as every 4th point in the interval
$1/Z \leq r \leq R_\text{cavity}$, where $R_\text{cavity}$  is the radius of
the cavity in which the $B$-splines basis orbitals are defined.
By cutting off points at short distances and using only every 4th
point in between we reduce the number of points by an order of magnitude.
Then, the Coulomb integrals are calculated in an extremely efficient way as
\begin{equation}
  q_k(jlmn) = \sum_{i=1}^{\mu}\rho_{jl}(r_i)Y_{kmn}(r_i)w_i.
\label{qk}
\end{equation}
Here $\mu \approx 100$ is number of points on the sub-grid and $w_i$ are
weight coefficients corresponding to a particular method of numerical
integration. Note that only one of two integrations for Coulomb
integrals is done on a reduced sub-grid. The initial integration (\ref{Yknm})
is done using all grid points.

\end{document}